\documentclass[pra, aps, twocolumn, superscriptaddress, nofootinbib, 10pt]{revtex4-2}

\usepackage{goldschmidt_aps}

\begin{document}

\title{Comparing and correcting robustness metrics for quantum optimal control}

\author{Andrew T. Kamen}
\affiliation{Pritzker School of Molecular Engineering, University of Chicago, Chicago, IL 60637}
\email{Corresponding author: andrewkamen@uchicago.edu}

\author{Samuel Fine}
\affiliation{Department of Computer Science, University of Chicago, Chicago, IL 60637}

\author{Bikrant Bhattacharyya}
\affiliation{California Institute of Technology, 1200 E. California Blvd., Pasadena, CA 91125}

\author{Frederic T. Chong}
\affiliation{Department of Computer Science, University of Chicago, Chicago, IL 60637}

\author{Andy J. Goldschmidt}
\affiliation{Johns Hopkins Applied Physics Laboratory, Laurel, Maryland 20723, USA}

\thispagestyle{plain}
\pagestyle{plain}

\begin{abstract}
Control pulses that nominally optimize fidelity are sensitive to routine hardware drift and modeling errors.
Robust quantum optimal control seeks error-insensitive control pulses that maintain fidelity thresholds and obey hardware constraints.
Distinct numerical approximations to the first-order error susceptibility include adjoint end-point and toggling-frame approaches.
Although theoretically equivalent, we provide a novel, systematic study demonstrating important numerical differences between these two approaches.
We also introduce a critical discretization correction to the widely-used toggling-frame robustness estimator, measurably improving its estimate of first-order error susceptibility.
We accomplish our study by positioning robustness as a first-class objective within direct, constrained optimal control.
Our approach uniquely handles control and fidelity constraints while cleanly isolating robustness for dedicated optimization.
In both single- and two-qubit examples under realistic constraints, our approach provides an analytic edge for obtaining precise, physics-informed robustness. 
\end{abstract}

\maketitle

 \begin{figure}[t]
    \centering
    \includegraphics[width=\columnwidth]{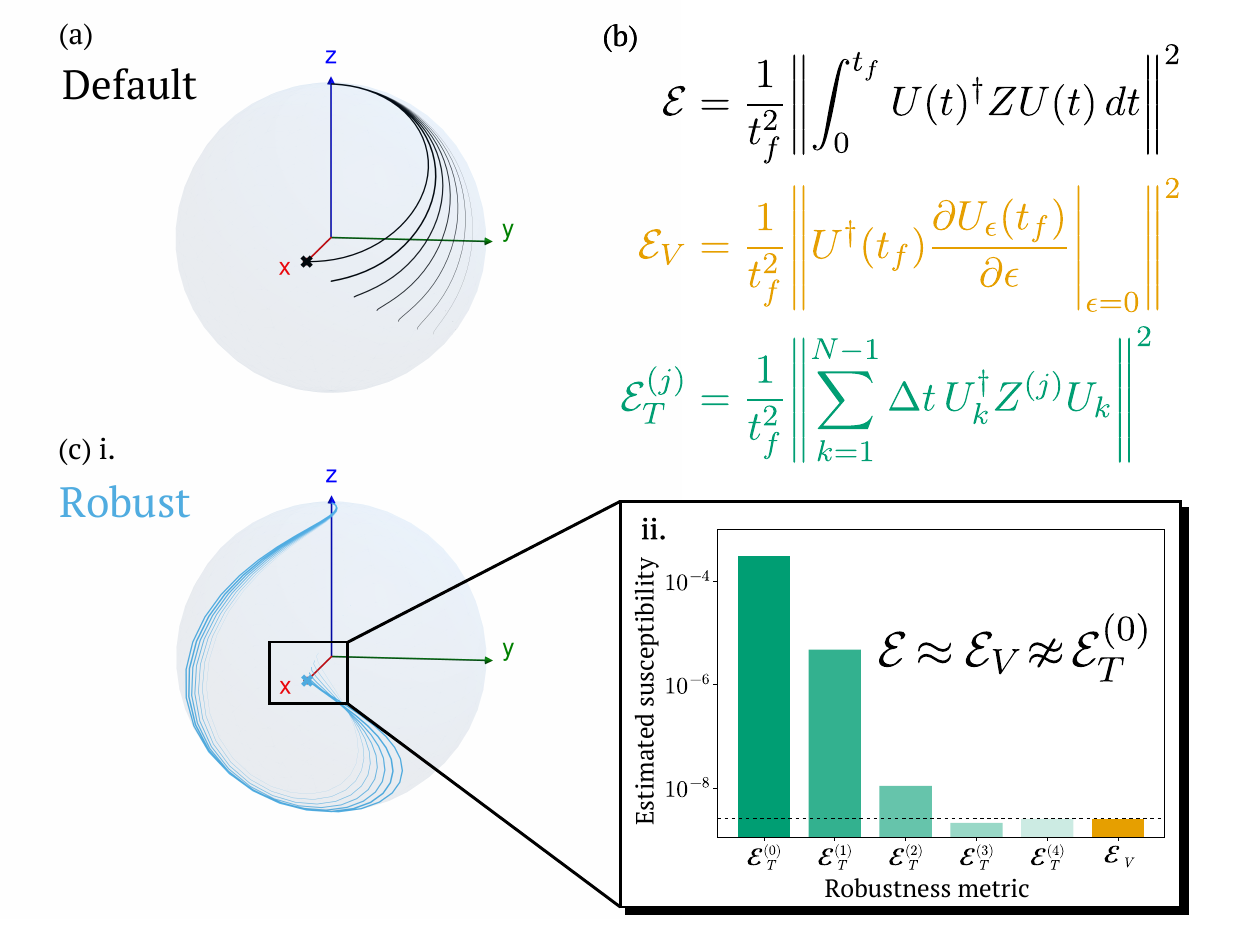}
    \caption{(a)~Qubit trajectory for a Hadamard gate without robust control. The lighter shades show deviations as the coherent error, $\epsilon Z$, increases. (b)~Robustness metrics [Sec.~\ref{sec:robustness}]: $\mathcal{E}$ is the first-order error susceptibility, $\mathcal{E}_V$ is the \textit{adjoint} objective, $\mathcal{E}_T^{(j)}$ is the (corrected) \textit{toggling} objective.  
    (c)~i.~Robust control of a Hadamard gate under the same coherent error as~(a). ii.~The bar chart compares the different robustness metrics. Ideal $\mathcal{E}$ (dashed line) is met by the adjoint objective (orange), but the toggling objective (green) requires a fourth order correction,  $\mathcal{E}_T^{(4)}$.
    }
    \label{fig:main-figure}
\end{figure}

\section{Introduction} \label{sec:introduction}
Fault-tolerant quantum computing requires high-fidelity quantum gates to satisfy the threshold theorem~\cite{nielsen2010quantum}. Often, quantum gates are implemented using Gaussian control waveforms that traverse short, geodesic paths in Hilbert space~\cite{nielsen2006quantum}. Gaussian controls are fast and straightforward to calibrate, but they are also highly sensitive to coherent errors from, e.g., hardware drift, miscalibration, or crosstalk~\cite{krantz2019quantum}. These errors kick the quantum state off its intended path [Fig.~\ref{fig:main-figure}~(a)]. Robustness to common coherent errors is a critical feature of useful quantum controls, and applications are ubiquitous across hardware platforms~\cite{koch2022quantum}, e.g., superconducting qubits~\cite{viola1998dynamical,motzoi2009simple,carvalho2021error,zhou2023quantum}, ions~\cite{mount2015error,leung2018robust}, atoms~\cite{evered2023high,hu2025universal}, and spins~\cite{analytic_spin_robustness2023, robust_spin_Ansel_2021}. Robust control improves the performance baseline and takes pressure off other parts of the quantum computing stack, such as the resource overhead for quantum error correction~\cite{nielsen2010quantum}. Lower sensitivity to drifting parameters also reduces recalibration frequency, resulting in longer device uptimes. Beyond quantum computing, the inverse of robustness can be optimized to improve quantum sensors~\cite{poggiali2018, haberle2013, rembold2020introduction}.

The price of robustness is control complexity. Robust controls must induce non-standard paths through Hilbert space in order to average out the impact that errors have on the dynamics~\cite{zeng2025fundamental}. Complicated controls compete against hardware constraints, calibration ability, decoherence at longer durations, and crosstalk or leakage due to wider control bandwidths~\cite{koch2022quantum}. Managing these trade-offs is critical for designing state-of-the-art robust control pulses.

Optimal control is one way to pursue robustness. Modern optimal control solves trajectory optimization problems: states and controls are discretized over time into knot points and an objective function of these decision variables is minimized, subject to various constraints. There are two broad classes of trajectory optimization: {(1)}~indirect optimization, which treats controls as the decision variables and enforces the dynamics implicitly by integrating the corresponding differential equation, and {(2)}~ direct optimization, which treats both states and controls as decision variables and enforces dynamics explicitly via a constraint between states and controls. Examples of the former are Gradient Ascent Pulse Engineering (GRAPE)~\cite{grape}, Chopped RAndom Basis (CRAB)~\cite{crab}, and Krotov's Method~\cite{goerz2019krotov}. Examples of the latter include Piccolo~\cite{trowbridge2023direct} and Ref.~\cite{petersson2025time}. Unlike indirect optimization, direct optimization efficiently supports general nonlinear constraints on the states and controls, which is useful for managing important nonlinear constraints like fidelity thresholds~\cite{trowbridge2023direct} or control bandwidth bounds~\cite{oda2023optimally}.

In quantum optimal control, robustness is usually implemented by penalizing a first-order error susceptibility~\cite{poggi2024universally, propson2022robust}; alternatives like sampling-based approaches scale unfavorably~\cite{dong2015sampling, wright2025superconducting}. As systems (and problem sizes) grow, it is essential that the first-order susceptibility can be computed efficiently in a manner that accurately realizes the desired robustness on hardware. Different numerical approaches exist---from \textit{toggling}-frame objectives~\cite{poggi2024universally} to \textit{adjoint}-state approaches~\cite{propson2022robust}---but the equivalence of their numerical performance and their accurate representation of the desired first-order susceptibility are usually taken for granted. To the best of our knowledge, we provide the first systematic comparison of these metrics and elucidate some important, practical corrections accounting for numerical discretization.

Our efforts are enabled by a framework for designing robust, quantum control signals using direct, constrained trajectory optimization. We demonstrate why each component--\textit{direct} and \textit{constrained}--is necessary for effective, systematic design of robust control signals. To support the easy adoption of our framework, we provide an extensible problem template for robust control within an open source software ecosystem for quantum optimal control~\cite{piccolo2025}.

Taken together, our novel contributions are as follows:
\begin{itemize}
    \item[$\circ$] We identify distinct performance gains of direct, constrained optimization over indirect optimization for design of constrained, robust quantum controls, illustrated with simple numerical examples.
    \item[$\circ$] We provide a comparative study of the primary methods for minimizing first-order sensitivities. This analysis introduces an important, perturbative correction that offsets the impact of discretizing a commonly-used robustness metric.
    \item[$\circ$] Our framework uniquely enables a systematic comparison of robust control designs. With simulations of one- and two-qubit gates, we quantify the benefit of physics-informed robust controls over universally-robust solutions.
\end{itemize}

We utilize the framework of direct, constrained trajectory optimization to systematically balance the many design trade-offs enforced by real-world systems. In contrast to other constrained approaches to robust control like space-curve geometry~\cite{spacecurves2023} or analytic methods~\cite{analytic_spin_robustness2023}, we emphasize the practical, numerical viability of our approach. Our proposed framework is flexible and scalable. We can transition between different hardware platforms and application-specific design goals by simply changing out the Hamiltonian.

The manuscript contains the following sections: 
Sec.~\ref{sec:to} reviews indirect (Sec.~\ref{sec:ito}) and direct (Sec.~\ref{sec:dto}) trajectory optimization. The contextualization within quantum optimal control follows (Sec.~\ref{sec:qoc}).
Sec.~\ref{sec:robustness} introduces robust control by way of the first-order error susceptibility. Different numerical approaches for calculating the first-order error susceptibility are discussed (Sec.~\ref{sec:toggling}-\ref{sec:adjoint}). Universal robustness (and its numerical calculation) is also introduced (Sec.~\ref{sec:universal}).
The remaining sections demonstrate the key contributions with numerical examples. First, Sec.~\ref{sec:comparing-constraints} compares direct and indirect optimization. The purpose is to explain why direct optimization is more appropriate for the nonlinear constraints on which the subsequent studies rely. Next, Sec.~\ref{sec:comparing-robustness} compares different robustness methods using constrained, direct trajectory optimization. We study differences among the robustness metrics via an interpretable Pareto frontier in the space of robustness versus control acceleration---all the while constrained within a practical fidelity threshold. Finally, Sec.~\ref{sec:practical-robustness} provides an application of our framework to the design of robust single and two-qubit gates that optimally balance hardware trade-offs. We argue that physics-informed design of robust protocols should be preferred in practice over universally-robust protocols.
Sec.~\ref{sec:conclusion} concludes with some recommendations for further efforts and suggestions for hardware studies.

\section{Trajectory optimization} \label{sec:to}

The evolution of a closed quantum system is governed by the time-dependent Schr\"odinger equation, $\partial_t \ket{\psi(t)} = -i H(t) \ket{\psi(t)}$, where $H(t)$ is the Hamiltonian generator and $\ket{\psi(t)}$ is the state. The solution is, formally, $U(t_f)=\mathcal{T}\exp\left(-i\int_0^{t_f}H(\tau)d\tau\right)$, where $\mathcal{T}$ is the time-ordering operator, $t_f$ is the final time, and $U$ is the unitary propagator, $\ket{\psi(t_f)} =U(t_f) \ket{\psi(0)}$. A controlled quantum system involves time-independent Hamiltonians $H_0$, which are static, and $H_j$ controlled linearly by functions $u_j(t)$---taken together, $H(\mathbf{u}(t)) = H_0 +\sum_{j=1}^{J} u_j(t) H_j$.

\textit{Trajectory optimization} is a categorical descriptor for the following problem:
\begin{align}
    \underset{\mathbf{x}(t),\, \mathbf{u}(t)}{\text{min}} &\quad \int_0^{t_f} \ell(\mathbf{x}(t), \mathbf{u}(t), t) \, dt  + \ell_{f}(\mathbf{x}(T)) \label{eq:to-prob} \\
    \text{s.t.} &\quad \dot{\mathbf{x}}(t) = -i H(\mathbf{u}(t)) \mathbf{x}(t) \label{eq:to-dynamics} \\
    &\quad \mathbf{x}(0) = \mathbf{x}_{\text{init}} \nonumber \\
    &\quad \mathbf{g}(\mathbf{x}(t), \mathbf{u}(t), t) \le 0, \label{eq:to-constraints}
\end{align}
where $\mathbf{x}(t)$ is the state trajectory (e.g., $\ket{\psi(t)}$, $U(t)$), $\mathbf{u}(t)$ is the control trajectory, $t \in [0, t_f]$, $\ell(\cdot, \cdot, \cdot)$ is an arbitrary objective (loss) function along the entire trajectory, and $\ell_f(\cdot)$ is a final objective. Eq.~\eqref{eq:to-constraints} lists $\mathbf{g}(\cdot, \cdot, \cdot)$ as a catch-all for additional constraints (examples in Sec.~\ref{sec:comparing-constraints}).  In general, Eq.~\eqref{eq:to-dynamics} can refer to any differential equation capturing the controlled dynamics---here, we have specialized to the Schr\"odinger equation.

In our work, we follow the \textit{discretize-then-optimize} approach: before solving Eq.~\eqref{eq:to-prob}, we first approximate the continuous dynamics by discretizing into $N= t_f / \Delta t$ points, such that $x_k = x(k \Delta t)$ and $u_k = u(k \Delta t)$. The collection of all decision variables at the iterate $k$ is called a \textit{knot point}. In our work, we will use $\mathbf{x}(t) := U(t)$ to design quantum gates. If we assume that controls are held constant between knot points---the \textit{zero-order hold} assumption---then the dynamics in Eq~\eqref{eq:to-dynamics} become $\mathbf{x}_{k+1} = \exp(-i H(\mathbf{u}_k) \Delta t) \mathbf{x}_k$.

\subsection{Indirect trajectory optimization} \label{sec:ito}

In \textit{indirect} trajectory optimization, we use the dynamics to eliminate the state variables in Eq.~\eqref{eq:to-prob}, such that
\begin{align} 
    \underset{\mathbf{u}_{1:N{-}1}}{\text{min}}& \quad \sum_{k=1}^{N-1} \ \ell_k(\mathbf{x}_k(\mathbf{u}_{1:k-1}), \mathbf{u}_k) + \ell_N(\mathbf{x}_N(\mathbf{u}_{1:N-1})) \label{eq:ito-prob} \\
    \text{s.t.} &\quad \mathbf{x}_0 = \mathbf{x}_{\text{init}} \nonumber \\
    &\quad \mathbf{g}_k(\mathbf{x}(\mathbf{u}_{1:k{-}1}), \mathbf{u}_k) \le 0. \label{eq:ito-constraints}
\end{align}
Notice that, e.g., $U(t_f) = \prod_{k=1}^{N-1} \exp(-iH(u_k)\Delta t)$ is a nonlinear function of the controls. This approach is sometimes referred to as a \textit{single-shooting} method because there is one evolution from a single initial condition. When possible, we want to leverage the temporal structure of Eq.~\eqref{eq:ito-prob}-\eqref{eq:ito-constraints} for efficiency~\cite{propson2022robust}. In the quantum optimal control literature, the hard constraint $\mathbf{g}$ is often removed (or scaled and added to the objective as a \textit{penalty}) to make Eq.~\eqref{eq:ito-prob} an unconstrained optimization over the landscape of controls. Unconstrained quantum control problems have favorable global convergence properties for fidelity objectives~\cite{rabitz2004quantum}, while constraints or modified objectives make landscapes more difficult (and interesting!) to navigate~\cite{riviello2015searching,larocca2022diagnosing}.

\subsection{Direct trajectory optimization} \label{sec:dto}

In \textit{direct} trajectory optimization, we retain both the controls and states,
\begin{align}
    \underset{\mathbf{x}_{1:N},\, \mathbf{u}_{1:N{-}1}}{\text{min}} &\quad \sum_{k=1}^{N-1} \ \ell_k(\mathbf{x}_k, \mathbf{u}_k) + \ell_N(\mathbf{x}_N) \label{eq:dto-prob} \\
    \text{s.t.} &\quad \mathbf{x}_{k+1} = \exp(-i H(\mathbf{u}_k) \Delta t) \mathbf{x}_k \label{eq:dto-dynamics} \\
    &\quad \mathbf{x}_0 = \mathbf{x}_{\text{init}} \nonumber \\
    &\quad \mathbf{g}_k(\mathbf{x}_k, \mathbf{u}_k) \le 0. \label{eq:dto-constraints}
\end{align}

This is a nonlinear program---to solve, we rely on generic nonlinear programming solvers like IPOPT~\cite{wachter2006ipopt}. The direct method uses constraints to enforce the dynamics between each pair of sequential knot points. Direct trajectory optimization is a \textit{multiple-shooting} method because each knot point is separately evolved for one time step~\cite{trowbridge2023direct, petersson2025time}. In this work, we rely specifically on \texttt{Piccolo.jl}, an open-source quantum control ecosystem written in Julia~\cite{piccolo2025}. We justify our use of direct trajectory optimization by comparing its performance with indirect optimization for representative problems with constraints in Sec.~\ref{sec:comparing-constraints}.

\subsection{Quantum optimal control} \label{sec:qoc}

In quantum optimal control, the form of common objectives and constraints vary slightly depending on the state, $\mathbf{x}$, under consideration. For closed systems, the state can be the vector or the unitary respecting the Schr\"odinger equation. For open systems, the state can be a density matrix or a quantum channel respecting the corresponding Lindblad master equation. For simplicity, we assume that the state, $\mathbf{x}$, is the unitary matrix, $U \in \mathbb{C}^{d \times d}$, and the dynamics are closed, such that the Schr\"odinger equation is obeyed. 

The most common objective for a unitary trajectory is to prepare a goal unitary, $G$, starting from the identity, $I$. This problem class corresponds to implementing the fundamental gate operations of a quantum computer. The accuracy of the preparation is quantified by the \textit{unitary fidelity}, 
\begin{equation} \label{eq:fid}
    \mathcal{F} = \frac{1}{d}\left|\Tr\{U_N^\dagger G\}\right|,
\end{equation}
with $U_N$ is the unitary at the final knot point, $N$, and $\Tr\{U_N^\dagger G\}$ is the Hilbert-Schmidt inner product between the operators $U_N, G \in\mathbb{C}^{d \times d}$. 

\section{Robustness metrics} \label{sec:robustness}

Achieving high-fidelity control of quantum devices requires that operations are unaffected by extant noise or system imperfections. Non-robust optimal control can yield controls with nominally high fidelity that deteriorate by several orders of magnitude in the presence of typical noise channels~\cite{soare2014experimental,zhou2023quantum}. Rather than optimizing for unrealistic and idealized systems, robust methods seek controls that are reliably high-fidelity in spite of physical errors and uncertainty. 

This section uses time-dependent perturbation theory to formalize the notion of robustness in terms of error susceptibility. Three ways to numerically quantify error susceptibility are outlined: the toggling, universal, and adjoint approaches. 

In what follows, we use the Frobenius (two-norm) of an operator, which is defined using the Hilbert-Schmidt inner product: $\norm{A}^2 = \frac{1}{d}\left|\Tr\{A^\dag A\}\right|$. 

\subsection{First-order susceptibility}

Suppose that the quantum dynamics, $i \partial_t U(t) = H(t) U(t)$, experience an error, $E(t)$. The impact of a small perturbation $H(t) + \epsilon  E(t)$ on the intended dynamics $U(t)$ can be studied using the language of time-dependent perturbation theory~\cite{poggi2024universally, zeng2019geometric, buterakos2021geometrical}. Let $\partial_t U_\epsilon (t) = -i(H(t) + \epsilon E(t)) U_\epsilon (t)$. Going to the interaction picture, write $U_\epsilon (t) := U(t) \widetilde{U}(t)$, where $U(t) = U_0(t)$, $\partial_t \widetilde{U}(t) = -i \epsilon  \widetilde{H}(t) \widetilde{U}(t)$ with $\widetilde{H}(t) = U^\dagger(t) E(t) U(t)$ (see Appendix~\ref{apdx:interaction}). The solution to this differential equation for the interaction picture propagator (setting $\widetilde{U}(0) = I$) is a time-ordered integral that can be approximated by a truncated Dyson series,
\begin{align}\label{eq:dyson}
    \widetilde{U}(t) &= \mathcal{T} \exp\left(-i \int_0^{t_f} \epsilon \widetilde{H}(t)\, dt \right) \\
           &\approx I - i \epsilon \int_0^{t_f} \widetilde{H}(t)\, dt + \mathcal{O}(\epsilon^2).
\end{align}
Define the \textit{first-order susceptibility} to the error, $E(t)$, using the first-order term in the Dyson series:
\begin{equation} \label{eq:first-order-sus}
    \mathcal{E}(E(t)) := \frac{1}{t_f^2} \norm{ \int_0^{t_f}  U(t)^\dagger E(t) U(t) \, dt}^2,
\end{equation}
where a normalization has been taken over the duration of the operation (it can also be useful to non-dimensionalize with respect to $E$). Eq.~\eqref{eq:first-order-sus} can be used to make a quantum control problem robust to $E(t)$. Note that errors are often treated as quasi-static on the timescale of quantum operations. In such cases, $E(t) \rightarrow E$. To use Eq.~\eqref{eq:first-order-sus} in optimization problems, we first have to discretize it.

\subsubsection{Discretization: A toggling approach} \label{sec:toggling}

When we refer to the numerical computation of susceptibility via Eq.~\eqref{eq:first-order-sus}, we call it the \textit{toggling} method, $\mathcal{E}_T(\cdot) := \mathcal{E}(\cdot)$, in line with previous literature~\cite{toggle_name_ref}. The numerical approximation of the integral in Eq.~\eqref{eq:first-order-sus} can be accomplished perturbatively in the timestep, $\Delta t$, such that
\begin{equation} \label{eq:toggling}
     \mathcal{E}^{(j)}_T((E_k)_{k=1}^{N-1}) := \frac{1}{t_f^2} \norm{\sum_{k=1}^{N-1}  \Delta t \, U_k^\dagger E^{(j)}_k U_k }^2,
\end{equation}
where
\begin{align}
    E^{(j)}_k &:= \sum_{n=0}^{j} \frac{i^{n} \Delta t^{n}}{(n+1)!} \text{ad}_{H_k}^{n}(E_k) \nonumber \\
     &= E_k + \frac{i \Delta t}{2!} [H_k, E_k] + \mathcal{O}(\Delta t^2),
\end{align}
and $E_k := E(k \Delta t)$ (additional details are provided in Appendix~\ref{apdx:discretizations}). Notice that Eq.~\eqref{eq:toggling} is a convenient choice for direct trajectory optimization. The states $U_k$ are already available as decision variables. No additional computation is needed to acquire them.

Keeping just the zeroth-order approximation to Eq.~\eqref{eq:toggling} is a common choice in the quantum control literature dating back to NMR \cite{magnus1954exponential, haeberlen1968coherent}. The added assumption of quasi-static errors makes $E_k = E,\,\forall k$, such that the approximation is
\begin{equation} \label{eq:first-toggling}
     \mathcal{E}^{(0)}_T(E) = \frac{1}{t_f^2} \norm{ \sum_{k=1}^{N-1}  \Delta t\, U_k^\dagger E U_k }^2,
\end{equation}
Eq.~\eqref{eq:first-toggling} is used in, e.g., \cite{ma2020optimal, tabuchi2017design, watanabe2024zz}. We emphasize that Eq.~\eqref{eq:first-toggling} is a specific, leading-order-in-$\Delta t$ approximation to the true, first-order-in-$\epsilon$ susceptibility~[Eq.~\eqref{eq:first-order-sus}].

\subsubsection{Universally-robust controls} \label{sec:universal}

Ref.~\cite{poggi2024universally} observes that Eq.~\eqref{eq:first-order-sus} can be modified to obtain a universally-robust bound for any constant error source, $E$. After a column-major vectorization \cite{poggi2024universally},
\begin{align} \label{eq:universal-derivation}
    \mathcal{E}(E) &= \frac{1}{t_f^2} || \int_0^{t_f} (U(t) \otimes U(t)^*) \vec{E}\, dt ||^2 \\
    & \le \frac{1}{t_f^2} || \int_0^{t_f} (U(t) \otimes U(t)^*) \,dt||^2 || \vec{E} ||^2
\end{align}
Define 
\begin{equation} \label{eq:universal-sus}
    \mathcal{E}_U := \frac{1}{t_f^2}\norm{ \int_0^{t_f} U(t) \otimes U(t)^* \, dt}^2
\end{equation}
as the \textit{universal first-order susceptibility}. The bound is universal because $\norm{\mathcal{E}(E)}^2 / \norm{E}^2 \le \mathcal{E}_U $ for any bounded, static coherent error, $E$.

The approximation of the integral in Eq.~\eqref{eq:universal-sus} can also be accomplished perturbatively in the timestep, $\Delta t$:
\begin{equation} \label{eq:universal-toggling}
    \mathcal{E}^{(j)}_U = \frac{1}{t^2_f} \norm{ \sum_{k=1}^{N-1} I_k^{(j)} \Delta t \left( U_k \otimes U_k^* \right) }^2,
\end{equation}
with
\begin{align}
    I_k^{(j)} &:= \sum_{n=0}^{j} \frac{\Delta t^n}{(n+1)!} \left(-i H_k \oplus i H_k^* \right)^{n} \\
    &= I + \frac{\Delta t}{2} \left(-i H_k \oplus i H_k^* \right) + \mathcal{O}(\Delta t^2),
\end{align}
where $\oplus$ is the Kronecker sum of the generators (additional details are provided in Appendix~\ref{apdx:discretizations}).

Again, keeping the zeroth-order approximation, we arrive at a more familiar numerical form for the universal robustness,
\begin{equation}
    \mathcal{E}^{(0)}_U = \frac{1}{t_f^2}\norm{\sum_{k=1}^{N-1} \Delta t \left( U_k \otimes U_k^* \right)}^2.
\end{equation}

\subsubsection{An adjoint approach} \label{sec:adjoint}

The adjoint approach is an alternative, equivalent form to Eq.~\eqref{eq:first-order-sus} that has been used, e.g., to solve for robust single qubit gates~\cite{propson2022robust} and to improve quantum sensors~\cite{liu2017quantum,pang2017optimal}.

To take the adjoint approach, expand the propagator, $U_\epsilon$, in a Taylor series about the perturbation, $U_\epsilon (t_f) \approx U_0(t_f)\left(I + \epsilon U^\dagger(t_f) \partial_\epsilon U_\epsilon (t_f)|_{\epsilon=0}\right) + \mathcal{O}(\epsilon^2)$. Matching orders of the perturbation with the Dyson series~[Eq.~\eqref{eq:dyson}], we have
\begin{equation}
    U^\dagger(t_f) \frac{\partial U_\epsilon (t_f) }{\partial \epsilon }\Bigg|_{\epsilon=0} = -i\int_0^{t_f} \widetilde{H}(t)\, dt.
\end{equation}
As such,
\begin{equation} \label{eq:adjoint-method}
    \mathcal{E}_V\left(E(t)\right) := \frac{1}{t_f^2} \norm{U^\dagger(t_f) \frac{ \partial U_{\epsilon}(t_f)}{\partial \epsilon }\Bigg|_{\epsilon=0}}^2
\end{equation}
is an equivalent way to compute the first-order susceptibility: $\mathcal{E}_V(E(t)) = \mathcal{E}(E(t))$. 

We call Eq.~\eqref{eq:adjoint-method} the \textit{adjoint} (or \textit{variational}) approach because $\partial_\epsilon U_{\epsilon}(t)|_{\epsilon=0}$ follows an adjoint equation of motion that is driven by the error Hamiltonian and dynamics,
\begin{equation}
    i \frac{\partial}{\partial t} \frac{\partial U_{\epsilon}(t)}{\partial \epsilon} = H_\epsilon(t) \frac{\partial U_{\epsilon}(t)}{\partial \epsilon} + E(t) U_{\epsilon}(t),
\end{equation}
where $H_\epsilon(t) := H(t) + \epsilon E(t) $. This equation of motion can be discretized under zero-order hold of the controls, $\mathbf{u}(t)$, and the error, $E(t)$. The joint dynamics constraints [Eq.~\eqref{eq:dto-dynamics}] for the states and adjoints become
\begin{equation} \label{eq:zoh-adjoint-dynamics}
    \begin{bmatrix} 
    U_{\epsilon,\,k+1} \\ 
    \frac{\partial U_{\epsilon,\,k+1}}{\partial \epsilon}
    \end{bmatrix}  =
    \exp \left(-i \Delta t \begin{bmatrix}
    H_{\epsilon, k} & 0 \\
    E_k &  H_{\epsilon, k}
    \end{bmatrix}
    \right)
    \begin{bmatrix}
    U_{\epsilon,\,k} \\
    \frac{\partial U_{\epsilon, k}}{\partial \epsilon}
    \end{bmatrix},
\end{equation}
where $H_{\epsilon, k} := H(\mathbf{u}_k) + \epsilon E_k$. Notably, unlike Sec.~\ref{sec:toggling} and Sec.~\ref{sec:universal}, there is no need to further develop Eq.~\eqref{eq:adjoint-method} with an expansion in $\Delta t$. The dynamics in Eq.~\eqref{eq:zoh-adjoint-dynamics} are exact under the assumption of zero-order~hold.

\section{Device-aware constraints}

Robust controls should be practical for use within the quantum computing stack, satisfying hardware limitations and performance requirements of quantum applications. Constraints reshape the optimization landscape of quantum control problems; we explore this fact in Sections~\ref{sec:comparing-constraints}. 

In this work, we use constraints on control curvature and gate fidelity to meet hardware and performance constraints, respectively. Curvature bounds quantify the expressivity of the control hardware.  Indeed, curvature is closely connected to bandwidth~\cite{oda2023optimally}: if the available control amplitudes are bounded, and the control frequency is capped by $\omega_0$, then $\norm{\ddot{\mathbf{u}}(t)}_2 \le \omega_0^2 \norm{\mathbf{u(t)}}_2$---larger curvature requires more bandwidth. Gate fidelity [Eq.~\eqref{eq:fid}] is an important performance metric for a quantum computer. Fidelity thresholds must be met at the physical layer if quantum error correction is to improve error rates at the logical layer~\cite{nielsen2010quantum}. Setting explicit fidelity constraints allows the optimizer to maximize objective performance without violating the threshold.

\section{Comparative Analysis: Hadamard~gate }

We compare indirect and direct optimization (Sec.~\ref{sec:comparing-constraints}) and the different robustness objects (Sec.~\ref{sec:comparing-robustness}) via an interpretable numerical example. We design a single qubit gate Hadamard gate, 
\begin{align} \label{eq:hadarmard}
        H=
        \frac{1}{\sqrt{2}}\begin{bmatrix}
            1 & 1 \\
            1 & -1
        \end{bmatrix}
\end{align}
with the control Hamiltonian, 
\begin{align}
    H(t) &= u_x(t) X + u_y(t) Y + u_z(t) Z.
\end{align}
The objective is to minimize the susceptibility due to the error Hamiltonian, $E(t)=Z$. We use $\mathcal{E}$ for the susceptibility integral [Eq.~\eqref{eq:first-order-sus}],  $\mathcal{E}_T$ for the toggling objective [Eq.~\eqref{eq:toggling}], and $\mathcal{E}_V$ for the adjoint objective [Eq.~\eqref{eq:adjoint-method}].

\subsection{Comparing constrained trajectory optimizations} \label{sec:comparing-constraints}

In our examples, we rely on direct trajectory optimization. To justify this choice, this section compares the performance of direct and indirect trajectory optimization under two particular constraints: controls~(Sec.~\ref{sec:acceleration}) and fidelity~(Sec.~\ref{sec:fidelity}). We show that indirect optimization problems can struggle to navigate constrained optimization landscapes that direct problems can traverse, and we connect this to the available decision variables. Recall that the indirect approaches only optimize over controls, $\mathbf{u}_{1:N-1}$, while direct methods have access to controls and states, $(\mathbf{x}_{1:N}, \mathbf{u}_{1:N-1})$.

\subsubsection{Controls constraints} \label{sec:acceleration}

\begin{figure}[t]
    \centering
    \includegraphics[width=\columnwidth]{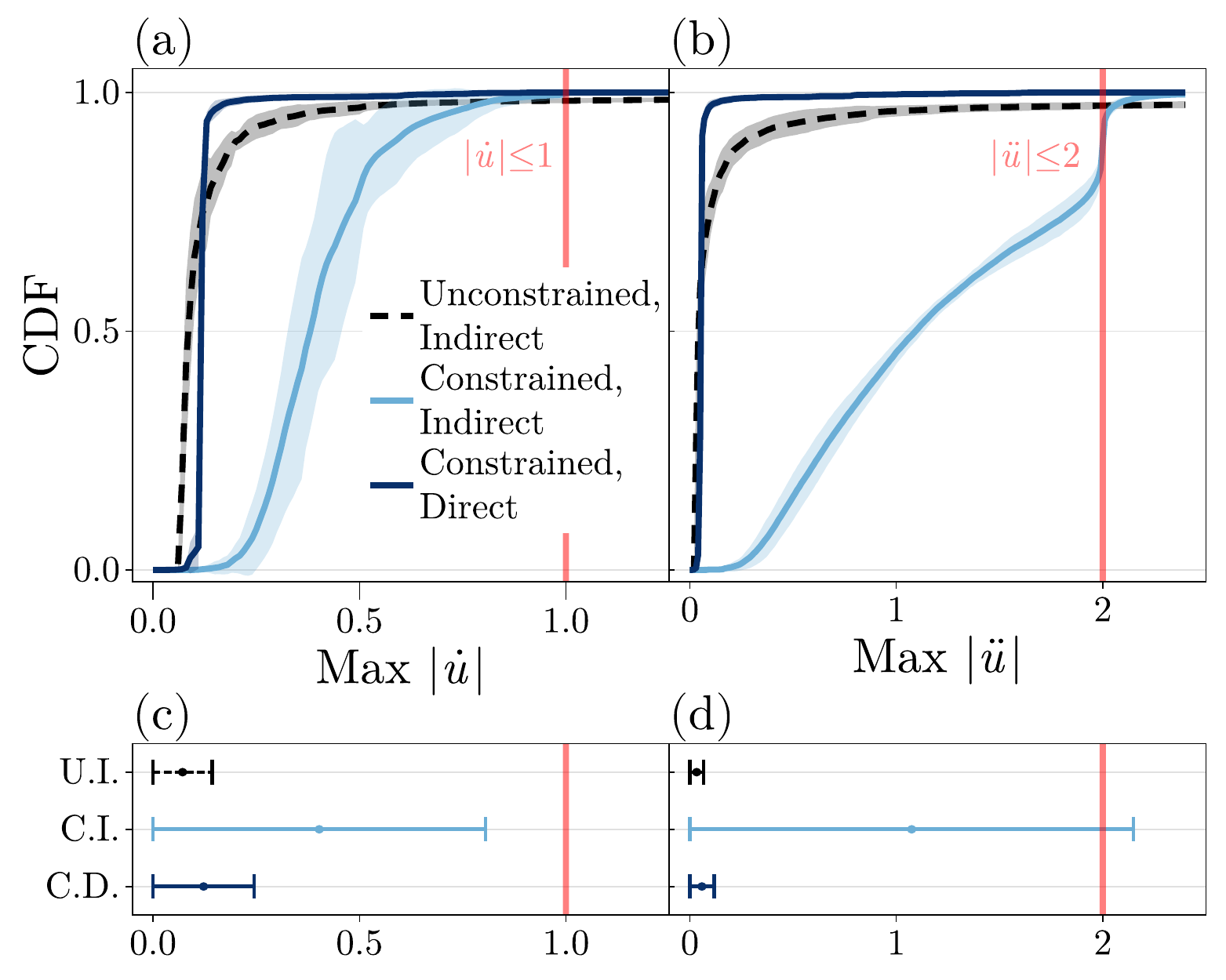}
    \caption{The cumulative distribution function (CDF) of the maximum control (a) velocity [(b) acceleration] over solver iterations for three trajectory optimization algorithms. The red vertical line marks the constraint. Constrained indirect (C.I.) spends more time near the boundary than constrained direct (C.D.), while the unconstrained indirect (U.I) optimization occasionally exceeds the boundary. At the final solver iteration, the maximum control (c) velocity [(d) acceleration] is below the constraint boundary.}
    \label{fig:CDF}
\end{figure}

\begin{figure}
    \centering
    \includegraphics[width=\columnwidth]{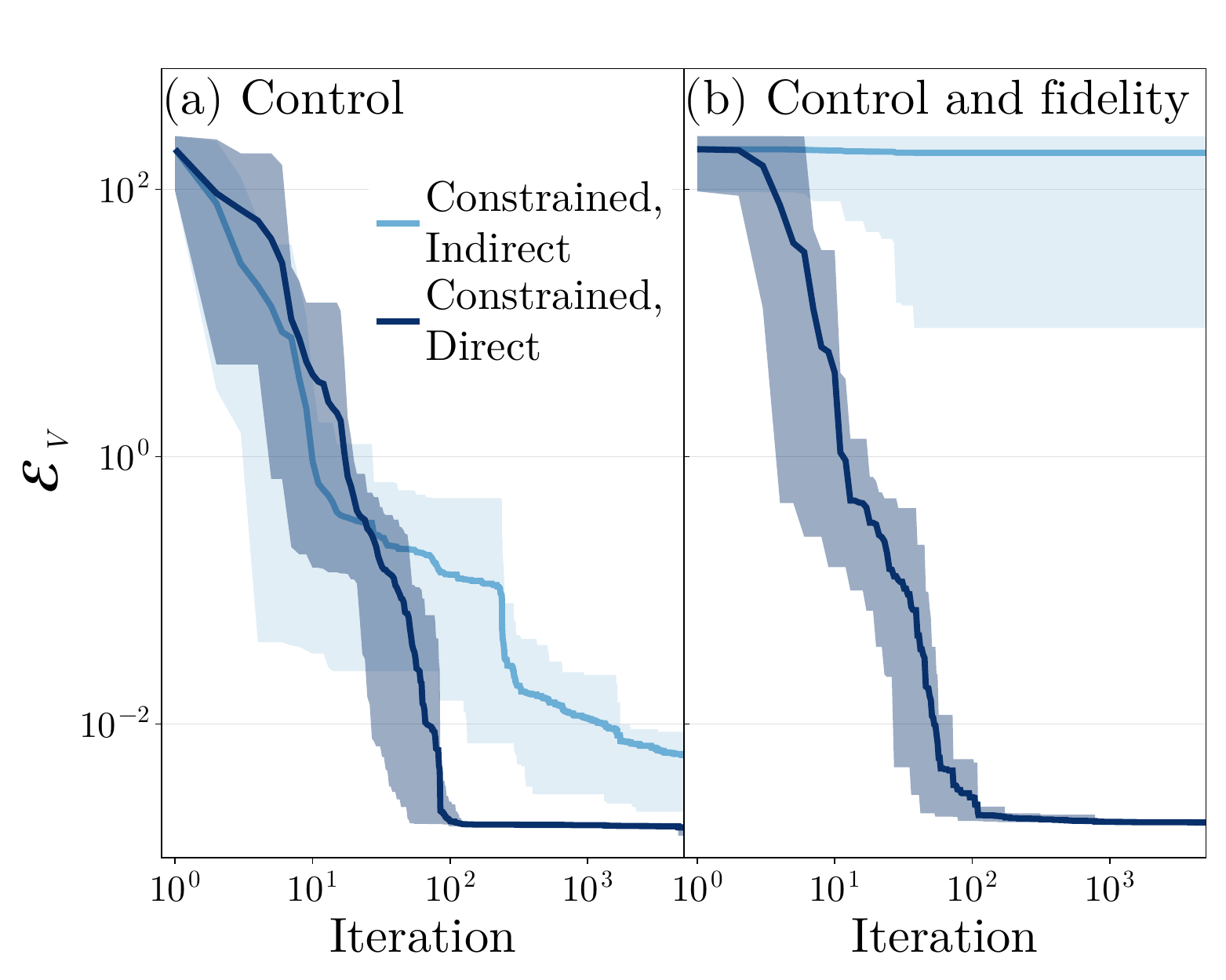}
    \caption{(a)~Solver convergence rates (measured using the adjoint objective, $\mathcal{E}_V$) are comparable for constrained indirect and constrained direct optimizers under control constraints. (b)~Adding a final fidelity constraint does not impact the convergence rate of the constrained direct optimizer, but the constrained indirect optimizer is unable to converge.}
    \label{fig:convergence}
\end{figure}

In Fig.~\ref{fig:CDF}, we compare the optimization path of indirect and direct optimization by way of the maximum velocity in \ref{fig:CDF}~(a), $\max_k \{\abs{\dot{u}_{x,k}}, \abs{\dot{u}_{y,k}}, \abs{\dot{u}_{z,k}} \}$ or acceleration in \ref{fig:CDF}~(b), $\max_{k} \{\abs{\ddot{u}_{x,k}}, \abs{\ddot{u}_{y,k}}, \abs{\ddot{u}_{z,k}} \}$. In particular, we plot the cumulative distribution function of the maximum control over each IPOPT iteration, which quantifies how often the optimization variables are close to the constraint during the solve. We set $\abs{\dot{u}} \le 1$ in (a) and $\abs{\ddot{u}} \le 2$ in (b); both constraint choices are large enough that at the final iteration, the value of the maximum control is below the bound in each of the three cases [Fig.~\ref{fig:CDF}~(c), (d)]. Fig.~\ref{fig:CDF} is the average performance across $10$ initial conditions with random controls.

Notice that the indirect, unconstrained optimization violates the control velocity and acceleration bound at a small percentage of iterations (the tail of the CDF exceeds the boundary); meanwhile, little-to-no population exceeds the boundaries during both constrained indirect and constrained direct optimizations, as expected. Similarities stop there. The constrained direct and constrained indirect cases take markedly different paths to solution. The direct optimizer stays well below the constraint throughout the solve. The indirect optimizer spends a lot of the time up against the boundary, as it attempts to follow the unconstrained indirect path until the constraints are activated.

For the most part, the differences between the implementation of constraints in direct and indirect are subtle. In indirect optimization [Eq.~\eqref{eq:ito-prob}], control constraints can be implemented by applying finite differences---once for velocity, twice for acceleration. Derivative bounds are linear constraints (n.b., to agree with Piccolo, the constraints in each case are slightly nonlinear because the timestep at each knot point is also a decision variable---useful for minimum time control~\cite{trowbridge2023direct}). In direct optimization [Eq.~\eqref{eq:dto-prob}], finite differences are also used to bound control velocity and acceleration, but an added trick can be introduced. The state variables can be augmented to include the control value, $u_k$, and the control velocity, $\dot{u}_k$, at each knot point $k$. State variables are available within indirect optimization. Hence, control acceleration, $\ddot{u}_k$ can be constrained to $\dot{u}_k$ via a single derivative constraint, and $\dot{u}_k$ can be constrained to $u_k$ in the same way. The trick reduces the complexity of the constraints by adding relatively few auxiliary decision variables.

Fig.~\ref{fig:convergence}~(a) shows via a plot of the robustness objective ($\mathcal{E}_V$) versus solver iteration that constrained indirect and constrained indirect methods have comparable convergence rates. The takeaway from Fig.~\ref{fig:CDF} should only be that qualitatively different paths are taken by unconstrained indirect, constrained indirect, and constrained direct optimizers---even if the final solutions are comparable. However, it also hints that indirect approaches uniquely experience added pressure from the control boundaries, which has greater implications in the next section.

\subsubsection{Fidelity constraints} \label{sec:fidelity}

In Fig.~\ref{fig:convergence}~(b), we constrain the final unitary fidelity to exceed $0.9999$. Observe that this new constraint does hurt the convergence of the indirect optimizer, which is now $1000\times$ less robust than the previous solution. There are stark differences between fidelity constraint implementations. In indirect optimization [Eq.~\eqref{eq:ito-prob}], states (unitary matrices, $U_k$) are not available as decision variables. Fidelity requires computing the final unitary, $U_N(u_{1:{N{-}1}})$, which is a nonlinear function of the controls via a dynamics rollout. The nonlinearity of the constraint increases with the number of knot points and the complexity of the numerical integration. In direct optimization [Eq.~\eqref{eq:dto-prob}], $U_N$ is a decision variable; therefore, the fidelity constraint depends overtly on $U_N$ via the Hilbert-Schmidt inner product [Eq.~\eqref{eq:fid}], making this essentially a linear constraint. The big, boundary-pushing adjustments to the controls during the indirect solve are unable to sufficiently inform the global, rollout-driven gradient of the fidelity constraint. In contrast, the small, distributed adjustments of the direct optimizer remain comfortably inside the control bounds while accommodating the fidelity constraint, enabling convergence.

\subsection{Comparing robustness metrics} \label{sec:comparing-robustness}

Having established in Sec.~\ref{sec:comparing-constraints} the importance of direct trajectory optimization [Eq.~\eqref{eq:dto-prob}] for handling nonlinear constraints like fidelity, in this section we rely on the direct approach to compare the different approaches to robustness. We move fidelity to a constraint, and optimize for one of our three robustness objectives plus a regularization on the smoothness of the controls: $Q \cdot \mathcal{E}_{T,U,V}(\mathbf{x}_{1:N}, \mathbf{u}_{1:N}) + R \cdot \mathcal{R}(\mathbf{u}_{1:N})$. The weights $Q$ and $R$ determine the shape of the cost landscape. Regularization, $\mathcal{R}(\mathbf{u}_{1:N}) = \sum_{j=1}^N \norm{\mathbf{u}_j}^2 + \norm{\dot{\mathbf{u}}_j}^2 + \norm{\ddot{\mathbf{u}}_j}^2$, induces smoothness in the controls~\cite{trowbridge2023direct}.

\begin{figure}[t]
    \centering
    \includegraphics[width=1.0\columnwidth]{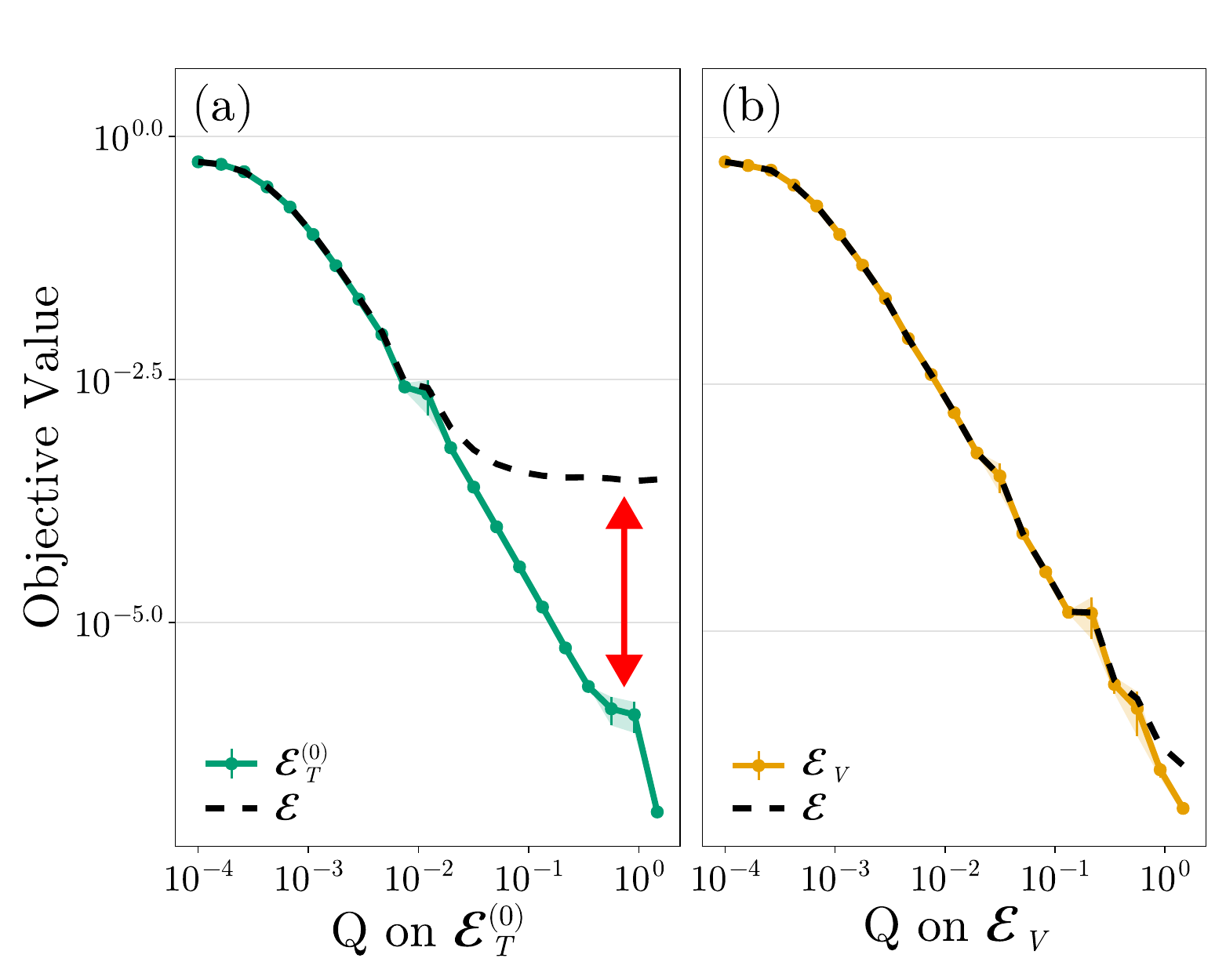}
    \caption{
    (a) The toggling objective [(b) adjoint objective] is optimized using a weight $Q$. The dashed black line is the first-order susceptibility of the optimized solution. 
    Increasing $Q$ prioritizes robustness over other regularization terms, so susceptibility should decrease. In (a), the desired first-order susceptibility saturates at large $Q$ because $\mathcal{E} \ne \mathcal{E}_T^{(0)}$.
    }
    \label{fig:comparison}
\end{figure}

In Fig.~\ref{fig:comparison} we use our direct optimizer to solve for a $32$~ns Hadamard while constraining the gate to a fidelity of $F\geq0.9999$. We optimize all single qubit controls $\mathbf{u} = [u_x, u_y, u_z]$ over $N=40$ knot points with spacing $\Delta t \coloneqq 0.8$. Each data point in Fig.~\ref{fig:comparison} is an average over $25$ random-control initializations.

\begin{figure}[t]
    \centering
    \includegraphics[width=1.0\columnwidth]{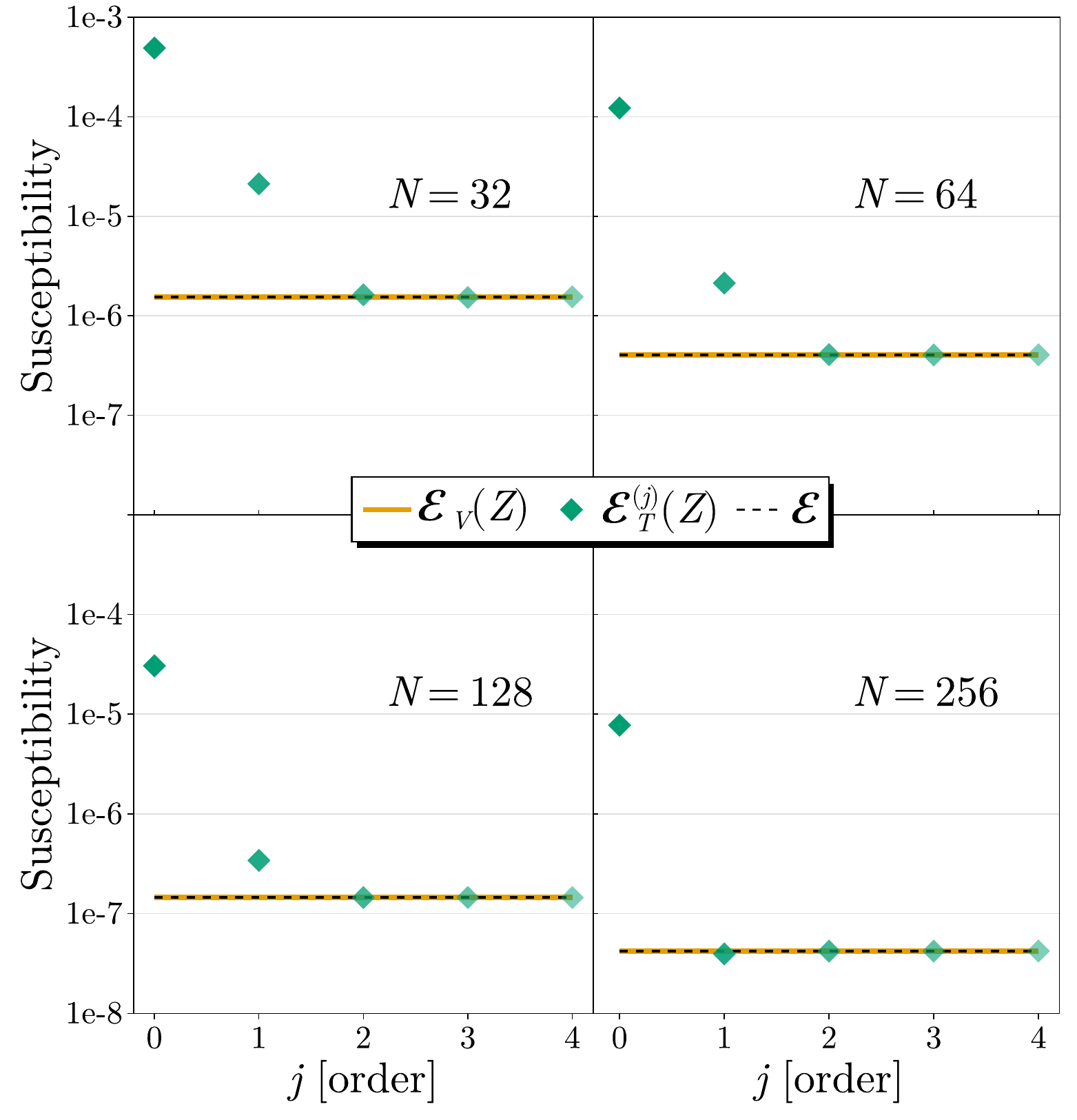}
    \caption{
    The optimizer minimizes the objective $\mathcal{E}_T^{(0)}$ for a $32$~ns Hadamard gate with an increasing number of knot points ($N$=32, 64, 128, 256). The black dashed line is the first-order susceptibility of the solution, $\mathcal{E}$. 
    The adjoint objective (orange line) is computed from the solution and matches $\mathcal{E}$, while the toggling objective (blue diamonds) only matches as the order $j$ in the $\Delta t$ expansion is increased. Higher orders are needed when $\Delta t$ is large ($N$ is small).
    }
    \label{fig:perturbation_theory}
\end{figure}

To optimize for the robust Hadamard, we minimize the toggling objective $\mathcal{E}_T^{(0)}$ and the variational objective, $\mathcal{E}_V$. Fig.~\ref{fig:comparison} shows us progressively increasing the robustness objective weight, $Q$, relative to the regularization, $R$. As a result, the optimizer more aggressively prioritizes reducing the objective value as we sweep. In comparing the two cases, we observe that the toggling objective decreases with increased weighting $Q$ in the optimization, but the actual susceptibility $\mathcal{E}$ saturates at around $Q \approx 10^{-2}$. The first order susceptibility $\mathcal{E}$ is an upsampled toggling objective with $2^{10}$ time steps over the duration of the gate. The time steps are sufficiently small, such that the Riemann sum approaches the actual value of the first-order susceptibility integral, but the number of knot points makes its immediate use in an optimizer infeasible.

The discrepancy between the toggling objective and the upsampled integral indicates that the optimizer is over-exploiting the discretization error in the toggling metric, minimizing the inaccurate metric rather than the actual susceptibility, $\mathcal{E}$. In contrast, the adjoint objective closely follows the upsampled objective, indicating the metric is accurate in the regime where $Q>0.01$, minimizing the correct objective even under aggressive robustness optimization.

In Figure \ref{fig:perturbation_theory} we demonstrate how the discretization error can be repaired with the perturbative expansion in orders of $\Delta t$. We see that increasing the order parameter $j$ results in the toggling objective approaching the accurate, upsampled integral ($2^{10}$ points, black dashed line); meanwhile, the adjoint objective agrees closely with the upsampled integral. Notably, higher orders are needed at smaller numbers of knot points (larger time steps), emphasizing that the toggling correction is particularly important when computational resource constraints motivate the use of fewer knot points.

\section{Practical robustness: \MakeLowercase{i}SWAP~gate} \label{sec:practical-robustness}

\begin{figure}[t]
    \centering
    \includegraphics[width=1.0\columnwidth]{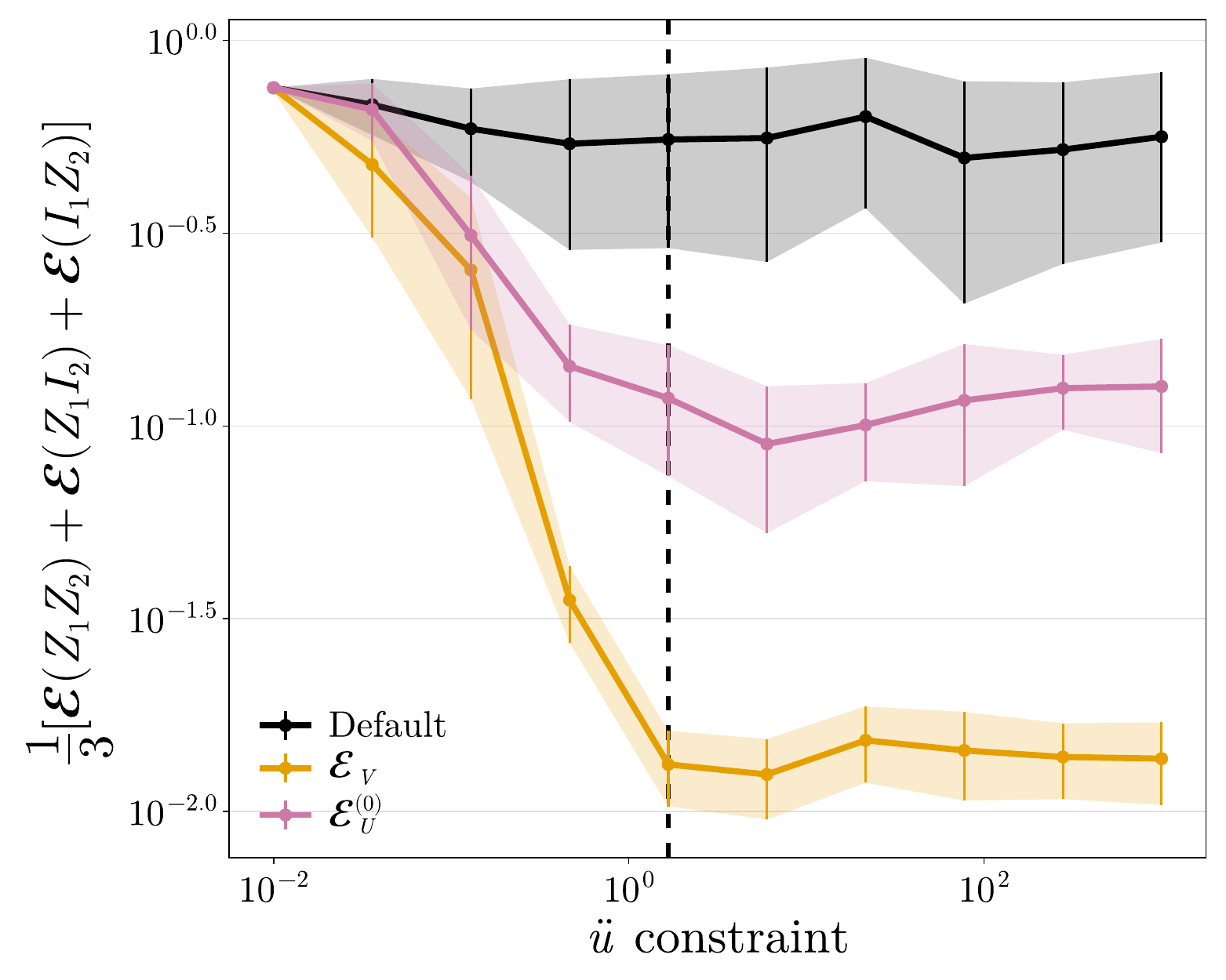}
    \caption{Robust two-qubit gate. The robustness of an iSWAP gate was optimized using three methods: minimizing gate infidelity without a robustness objective (default, black), minimizing $\mathcal{E}_V$ (adjoint, orange), and minimizing  $\mathcal{E}_U$ (Universal, pink). The fidelity is constrained to $\mathcal{F}\geq 0.9999$ and the control acceleration bound $\ddot{u}$ is gradually increased. The variational objective specifically targets $Z_1I_2$, $I_1Z_2$, and $Z_1 Z_2$ errors, converging to a solution that is more optimal for these specific errors when compared with the universal objective. 
    }
    \label{fig:dda}
\end{figure}

\begin{figure}[ht]
    \centering
    \includegraphics[width=1.0\columnwidth]{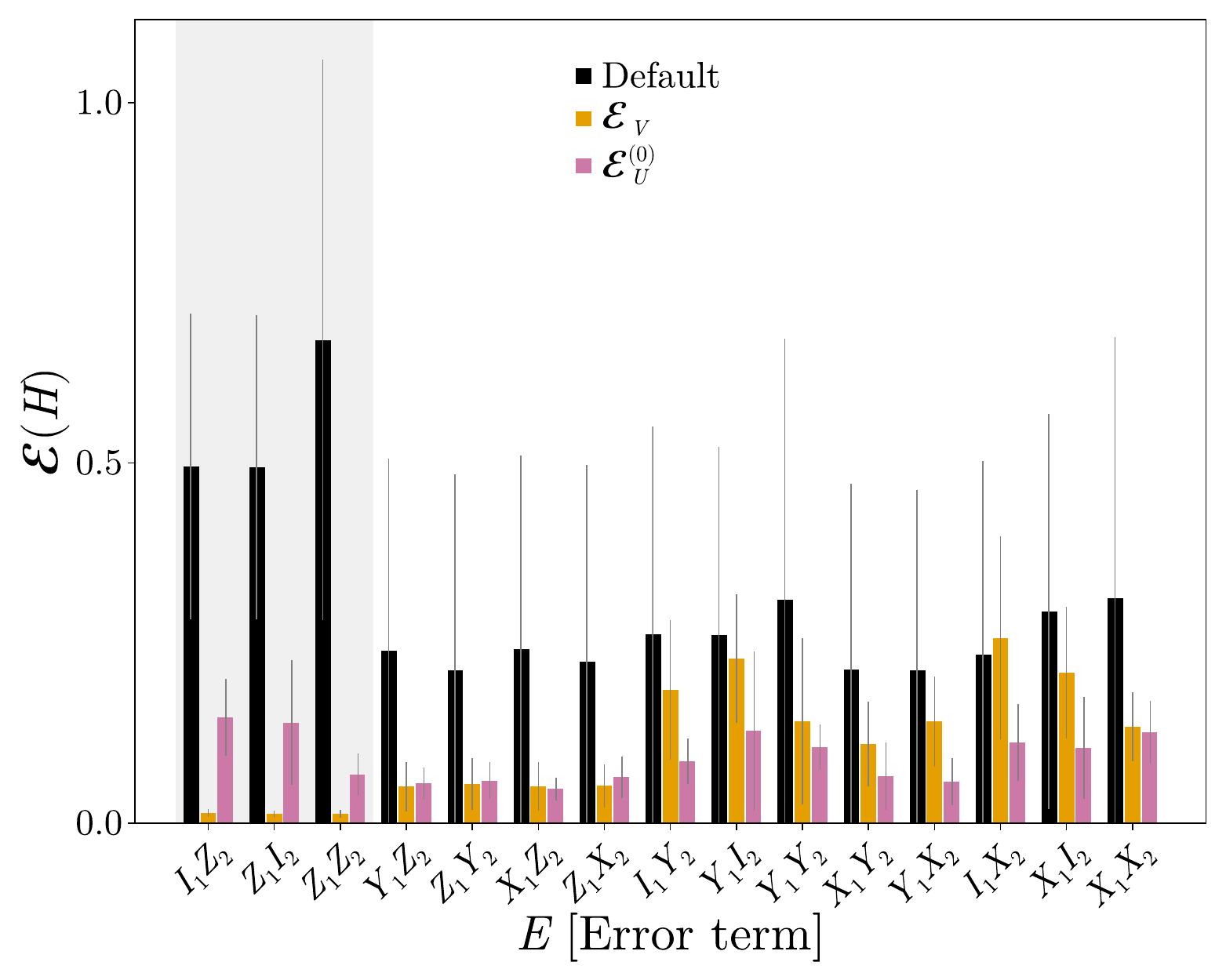}
    \caption{The error susceptibility to all 15 Pauli strings corresponding are plotted. The faint gray background is a guide for the eye, showing the error susceptibility for $Z_{(1)}$, $Z_{(2)}$, and $Z_{(1)} Z_{(2)}$ corresponding to the dominant error mechanisms in a two transmon system with tunable coupling. The targeted robustness outperforms the all-purpose objective.
    }
    \label{fig:suscept}
\end{figure}

As a physically-motivated example, we consider a system consisting of two transmons. Our goal is to design an iSWAP gate,
\begin{align}
        \text{iSWAP}=\exp{i\frac{\pi}{4}(X_1 X_2 + Y_1 Y_2)},
\end{align}    
with the following control Hamiltonian
\begin{equation}
    H(t) = \sum_{j=1,2} \vec{u}_j(t)\cdot\vec{P}_j + g(t) (X_1 X_2 + Y_1 Y_2)
\end{equation}
where $\vec{P}_j = \begin{bmatrix} X_j & Y_j & Z_j \end{bmatrix}$ is a vector of Pauli matrices, and subscripts $1$ and $2$ correspond to qubits $1$ and $2$, e.g., $X_1 = X \otimes I$. Physically, each transmon qubit is assumed to be controlled with microwave ($X$, $Y$) and flux ($Z$) drives. The $X_1 X_2 + Y_1 Y_2$ term corresponds to a transversal coupling between qubits with a tunable, time-dependent coupling, $g(t)$. The tunable coupling can be mediated by a virtual exchange interaction \cite{yan2018tunable} and has been implemented in \cite{mundada2019, wu2024modular, sung2021}. 

If we consider a system where the primary error channels are frequency shifts due to spectator qubits and parasitic $ZZ$ interaction, we can model the error Hamiltonian as
\begin{equation}
    E(t)= \epsilon_{1} Z_1 + \epsilon_{2} Z_2 + \epsilon_{12} Z_1 Z_2
\end{equation}
The above is often the dominant error channel in superconducting architectures with tunable coupling (e.g.,~\cite{cai2021impact, sung2021, tripathi2022suppression, zajac2021spectator, wu2024modular}). The first two terms $\epsilon_{1} Z_1$ and $\epsilon_{2} Z_2$ model noise from weakly coupled spectator qubits. Depending on the state of the spectator qubit, the frequency of the individual qubits will change due to a dispersive shift~\cite{cai2021impact, tripathi2022suppression, zajac2021spectator}. For the purposes of this study, we assume that the values of $\epsilon_1$ and $\epsilon_2$ are quasi-static, so the gate is much shorter than the time scales at which the state of the spectator qubits change. As a result, the frequency pulling by the spectator is assumed to be constant throughout the gate. Additionally, architectures with tunable coupling experience sensitivity due to parasitic $ZZ$ interactions, where the frequency of the two qubits participating in the gate give each-other frequency shifts depending on their individual states. In particular, Ref.~\cite{wu2024modular} characterizes these effects for an iSWAP gate between two resonant transmons, which is a fourth-order effect in Schrieffer-Wolff perturbation theory. For simplicity, we assume the parasitic $ZZ$ interaction strength is quasi-static. However, the non-linear and time-varying dependence of $\epsilon_{12}$ on $g(t)$ could be incorporated in a more detailed optimization problem, which we leave for future work. We note that such time-dependent errors for specific error channels can be incorporated in the adjoint equation of motion, but are incompatible with the universal objective, which remains agnostic to the type of error and is only applicable to quasi-static errors.

In Fig.~\ref{fig:dda}, we compare the adjoint, universal, and default methods by sweeping the constraint on the control acceleration. To ensure that the desired gate is achieved, we constrain the optimization to a fidelity of $\mathcal{F}=0.9999$. We compare the susceptibility using $\mathcal{E}_V$ as our metric (because it accurately reflects the true susceptibility) and run the optimization over $16$ random seeds. In Fig.~\ref{fig:dda}, we see typical Pareto frontiers, where increasing the upper bound on the control acceleration allows the optimizers to find solutions with lower susceptibility. Observe that the universal objective is unable to achieve comparable robustness for the dominant error channels. Under curvature and duration constraints, the universal objective must distribute its finite ability for robustness across all error operators, limiting its performance. This can be observed in Fig.~\ref{fig:suscept}, which shows the susceptibility of each error operator for the controls at the dashed line in Fig.~\ref{fig:dda}. Although the universal method outperforms the adjoint method for some error channels, the adjoint method provides significant improvements for the error channels of interest (gray background). We recommend the adjoint method whenever the set of Pauli strings corresponding to the dominant sources of error are known.

\section{Conclusion} \label{sec:conclusion}

We present an efficient software solution capable of optimizing objectives that accurately quantify robustness for quantum controls, while also satisfying the realistic constraints of quantum hardware and fault-tolerant quantum computation. The solution is integrated within an existing open-source ecosystem~\cite{piccolo2025}, enabling plug-and-play deployment across a wide range of quantum architectures and control tasks.

We compared the performance of direct and indirect trajectory optimization in constrained quantum control problems. Our results indicate that the direct approach has better convergence properties when realistic constraints on controls and fidelity are included.

We used the direct approach to re-examine the accuracy of standard metrics for robustness used in quantum optimal control and uncovered a critical discretization error affecting the performance of widely-used formulations of first-order susceptibility. Starting from time-dependent perturbation theory, we showed that the toggling, adjoint, and universal objectives are all derived from the same underlying expression; however, their discretizations hold critical discrepancies. In particular, the standard toggling objective and its universal counterpart misrepresent the first-order susceptibility because they assume the perturbed time evolutions are reasonably approximated to first-order in the length of each time step, $\Delta t$. We resolved this discrepancy by using the available optimization variables to recast the toggling metric into an optimizable expansion to arbitrary order in $\Delta t$. Furthermore, we noted that the adjoint method circumvents the discretization error by incorporating the perturbation into the adjoint equation of motion.

We demonstrated the practical implications of this error with numerical simulations. Notably, we found a point of saturation past which increasing the weight on the toggling objective in the optimizer did not further reduce the ground truth susceptibility, and instead led to over-fitting of the inaccurate metric. Using constrained optimization, we found that physics-informed robust control---in which errors are restricted to a physically motivated set of Pauli strings---is preferable to universal robustness. A natural extension is to move beyond the quasi-static perturbations considered here and to include time-dependent errors in the adjoint equation of motion. Using autocorrelation functions extracted from measured noise power spectral densities would allow the robustness objectives to reflect experimentally observed temporal correlations~\cite{soare2014experimental,oda2023optimally}.

To summarize, we discussed the advantages of implementing robust quantum control using direct rather than indirect trajectory optimization and analyzed the numerical accuracy of robustness metrics widely used in the quantum control community. In our theoretical and numerical analyses, we present an important correction to the discretization error that appears in both the toggling and universal robustness objectives. Finally, we showed that the adjoint approach provides an efficient and numerically accurate alternative for acquiring robust controls, offering a practical framework for implementing physics-informed robustness under realistic constraints.

\section*{Code availability}

\texttt{Piccolo.jl} is an open-source Julia package for quantum optimal control~\cite{piccolo2025}. We have added support for the adjoint, toggling, and universal control methods discussed in this paper.

\section*{Acknowledgments} \label{sec:ackn}
FTC is the Chief Scientist for Quantum Software at Infleqtion.

AJG was supported by an appointment to the Intelligence Community Postdoctoral Research Fellowship Program at University of Chicago administered by Oak Ridge Institute for Science and Education (ORISE) through an interagency agreement between the U.S. Department of Energy and the Office of the Director of National Intelligence (ODNI). 

This work is funded in part by the STAQ project under award NSF Phy-232580; in part by the US Department of Energy Office of Advanced Scientific Computing Research, Accelerated Research for Quantum Computing Program; and in part by the NSF Quantum Leap Challenge Institute for Hybrid Quantum Architectures and Networks (NSF Award 2016136), in part by the NSF National Virtual Quantum Laboratory program, in part based upon work supported by the U.S. Department of Energy, Office of Science, National Quantum Information Science Research Centers, and in part by the Army Research Office under Grant Number W911NF-23-1-0077. The views and conclusions contained in this document are those of the authors and should not be interpreted as representing the official policies, either expressed or implied, of the U.S. Government. The U.S. Government is authorized to reproduce and distribute reprints for Government purposes notwithstanding any copyright notation herein.

\appendix
\numberwithin{equation}{section}
\numberwithin{figure}{section}

\section{Interaction Hamiltonian} \label{apdx:interaction}

Let $\partial_t U_\epsilon (t) = -i(H(t) + \epsilon E(t)) U_\epsilon (t)$. In the interaction picture, recall that $U_\epsilon(t) =: U(t)\tilde{U}(t)$ defines the interaction-picture propagator, $\tilde{U}$. By the chain rule, $i\partial_t {U}_\epsilon = i\partial_t U \tilde{U} + U i \partial_t \tilde{U}$. Inserting the Schr\"odinger equation,
\begin{equation}
    (H + \epsilon E) U \tilde{U} = H U \tilde{U} + U i\partial_t \tilde{U}.
\end{equation}
Hence, the generator, $\tilde{H}$, is given by
\begin{equation}
    i \partial_t \tilde{U} = (U^\dagger E U) \tilde{U} =: \tilde{H} \tilde{U}.
\end{equation}

\section{Susceptibility Discretizations} \label{apdx:discretizations}

Discretizations in $\Delta t$ are performed in Eq.~\eqref{eq:toggling} and Eq.~\eqref{eq:universal-toggling} of the main text. Notably, these discretizations appear in powers of $\Delta t \cdot H_k$---for many systems, $H_k = H_0 + \sum_j a_{j,k} H_j$, and so our expansion tracks powers of $\Delta t \cdot a_\text{bound}$. Unless this quantity is $\ll 1$, the higher-order terms are quite relevant.

\subsection{Toggling} Assuming zero-order hold on controls and errors, the objective in Eq~\eqref{eq:first-order-sus} can be written
\begin{equation}
 \mathcal{E}_T = \frac{1}{t_f^2}\norm{\sum_{k=1}^{N-1} U_k^\dagger \left( \int_0^{\Delta t} e^{i H_k t} E_k e^{-i H_k t}\, dt\right) U_k}^2.
\end{equation}
Applying the Hadamard identity (a special case of Baker-Campbell-Hausdorff)~\cite{sakurai2017modern}, we have
\begin{align}
     \int_0^{\Delta t} e^{i H_k t} E_k e^{-i H_k t}\, dt 
     &= \Delta t \sum_{n=0}^\infty \frac{i^{n} \Delta t^{n}}{(n+1)!} \text{ad}_{H_k}^{n}(E_k)  \\
     &= \Delta t E_k + \frac{i \Delta t^2 }{2!} [H_k, E_k] + \mathcal{O}(\Delta t^2),\nonumber 
\end{align}
where $\text{ad}_H(\cdot)=[H,\cdot]$ is used to define the iterated commutator, $\text{ad}^n_H(\cdot)=[H,[H, \cdots[H,\cdot]]]$. These two identities provide us with Eq~\eqref{eq:toggling} in the main text.

\subsection{Universal} Again, assuming zero-order hold on controls, the objective in Eq.~\eqref{eq:universal-sus} can be written
\begin{equation}
    \mathcal{E}_U = \frac{1}{t_f^2}\norm{\sum_{k=1}^{N-1} \left( \int_0^{\Delta t} e^{\left(-i H_k \oplus i H_k^*\right) t} \, dt \right) \left( U_k \otimes U_k^* \right)}^2,
\end{equation}
where $\oplus$ denotes the Kronecker sum. Recalling $\varphi_1(z) := (e^z -1) / z$ familiar from the study of exponential integrators, we have
\begin{align}
    \int_0^{\Delta t} e^{L_k t} \, dt &= \Delta t\, \varphi_1( \Delta t L_k)  \\
    &= \Delta t\, I  + \frac{\Delta t^2}{2} L_k + \mathcal{O}(\Delta t^3). \nonumber
\end{align}
Taken together, we obtain Eq.~\eqref{eq:universal-toggling} in the main text.
\vfill

\bibliographystyle{apsrev4-2}
\bibliography{bibs/robust, bibs/preprints}

\end{document}